\documentclass[11pt]{article}

\usepackage{amssymb}

\usepackage{graphicx}

\usepackage[english,francais]{babel}

\newfont{\ensmathquatorze}{msbm10 scaled 1400}
\newfont{\ensmathonze}{msbm10 scaled 1100}
\newfont{\ensmathdix}{msbm10}
\newfont{\ensmathneuf}{msbm10 scaled 833}
\newfont{\ensmathhuit}{msbm10 scaled 694}
\newfam\ensmathfam                        
\textfont\ensmathfam=\ensmathonze        
\scriptfont\ensmathfam=\ensmathdix       
\scriptscriptfont\ensmathfam=\ensmathhuit
\def\ensmf{\fam\ensmathfam\ensmathonze}         

\def\be{\begin{equation}}
\def\ee{\end{equation}}

\def\bea{\begin{eqnarray}}
\def\eea{\end{eqnarray}}
\def\beann{\begin{eqnarray*}}
\def\eeann{\end{eqnarray*}}

\renewcommand{\leq}{\leqslant}
\renewcommand{\geq}{\geqslant}

\newcommand{\ket}[1]{|\kern.3ex#1\kern.3ex\rangle}
\newcommand{\bra}[1]{\langle\kern.3ex #1 \kern.3ex|}
\newcommand{\APPROX}[1]{                
   {{\raisebox{-.3cm}{$\textstyle\simeq$}} \atop {\scriptstyle{#1}}}}

\newcommand{\leadto}[1]{                
   {{\raisebox{-.3cm}{$\textstyle\longrightarrow$}} \atop {\scriptstyle{#1}}}}

\newcommand{\EXP}[1]{{\mbox{\large e}}^{#1}}         

\newcommand{\tr}[1]{\mathop{\mathrm{Tr}}\nolimits\left\{ #1 \right\}}  
\newcommand{\cotg}{\mathop{\mathrm{cotg}}\nolimits}  

\def\NN{{\ensmf N}}                 
\def\ZZ{{\ensmf Z}}                 
\def\QQ{{\ensmf Q}}                 
\def\RR{{\ensmf R}}                 

\def\I{{\rm i}}                  
\def\D{{\rm d}}                  
\def\Dc{{\rm D}}                 

\newcommand\ab{{\alpha\beta}}
\newcommand\ba{{\beta\alpha}}
\newcommand\mb{{\mu  \beta}}
\newcommand\bm{{\beta\mu}}
\newcommand\lab{l_{\alpha\beta}}

\begin{document}

\selectlanguage{english}

\title{Scattering theory on graphs (2)~:\\
       the Friedel sum rule}
 
\author{    Christophe Texier}

\date{22nd February 2002} 

\maketitle	

\noindent
{\small
Laboratoire de Physique Th\'eorique et Mod\`eles Statistiques.
Universit\'e Paris-Sud, B\^at. 100,\\
Laboratoire de Physique des Solides. 
Universit\'e Paris-Sud, B\^at. 510, F-91405 Orsay Cedex, France.
}

\vspace{0.5cm}

\noindent
E-mail~: 
\begin{minipage}[t]{7cm}
texier@ipno.in2p3.fr, texier@lps.u-psud.fr
\end{minipage}

\vspace{0.5cm}

\begin{abstract}
We consider the Friedel sum rule in the context of the scattering theory 
for the Schr\"odinger operator $-\Dc_x^2+V(x)$ on graphs made of 
one-dimensional wires connected to external leads.
We generalize the Smith formula for graphs.
We give several examples of graphs where the state counting method given by
the Friedel sum rule is not working. 
The reason for the failure of the Friedel sum rule to count the states is 
the existence of states localized in the graph and not coupled to the leads,
which occurs if the spectrum is degenerate and the number of leads too small.
\end{abstract}

\noindent
PACS~: 03.65.Nk, 72.10.Bg, 73.23.-b





\section{Introduction\label{sec:Intro}}

This article follows \cite{TexMon01} in which we have considered the 
scattering problem for the Schr\"odinger operator on graphs. The graphs
we are interested in are networks made of one-dimensional wires identified
with finite intervals of $\RR$ and being connected at vertices. 
The study of such systems has been shown to be relevant in many contexts
(for references see  
\cite{AvrSad91,Exn97,KotSmi99,KosSch99,PasMon98,AkkComDesMonTex00,TexMon01}).
For example graphs have been often considered as simple modelizations for the
mesoscopic networks realized experimentally. 
In this context the scattering theory is a fundamental tool involved 
in the study of transport properties and many other questions.
Several works have been devoted to the study of scattering theory on 
graphs among which we can quote \cite{GerPav88,Ada92,Exn97,TexMon01}.
In our work we examine an important aspect of scattering theory, namely 
the relation between the scattering and spectral properties that is 
established through the Friedel sum rule (FSR). The essence of the FSR is 
to count the number of states in the scattering region, that is related to 
the phases of the eigenvalues of the scattering matrix.
The purpose of this article is to show that one must be careful when
applying the Friedel sum rule to graphs since this formula does not hold
for any graph, one of the reasons of the breakdown being the occurence of
degeneracies in the spectrum of the graph (this is a necessary but not 
a sufficient condition as we will see).

Here we consider the Schr\"odinger operator
\be
H=-\Dc_x^2+V(x)
\ee
where $\Dc_x=\D_x-\I A(x)$ is the covariant derivative~; the $x$ coordinate
lives on the graph ${\cal G}$. We briefly recall the notations adopted in
previous works \cite{AkkComDesMonTex00,Des00,TexMon01}.
The graph ${\cal G}$ is made of $B$ one-dimensional wires connected at
$V$ vertices.
We will denote the vertices with greek letters ($\alpha$, $\beta$,
$\mu$,\ldots). The $V\times V$-adjacency matrix $a_\ab$ (or connectivity
matrix) is defined as: $a_\ab=1$ if the vertices $\alpha$ and $\beta$
are connected by a bond and $a_\ab=0$ otherwise. The coordination of the
vertex $\alpha$ (number of bonds issuing from the vertex) is therefore
$m_\alpha=\sum_\beta a_\ab$.
The bond between the vertices $\alpha$ and $\beta$ will be designated with
parenthesis: $(\ab)$.
We also introduce the notion of {\it arc} which is an {\it oriented bond}. 
Each bond $(\ab)$ is associated with two arcs: $\ab$ and $\ba$. The arcs 
are labelled with roman letters ($i$, $j$,\ldots ) and we
denote the reversed arc of $i$ with a bar: $\bar i$.

The coordinate $x_\ab$ on the bond $(\ab)$ of length $\lab$ belongs to the
interval: $x_\ab\in[0,\lab]$ where $l_\ab$ is the length of the bond $(\ab)$
(note that by definition $x_\ba=\lab-x_\ab$).

The Schr\"odinger operator acts on scalar functions $\psi(x)$ living on  
${\cal G}$ that are represented by a set of $B$ components 
$\psi_{(\ab)}(x_\ab)$ satisfying appropriate boundary conditions at the 
vertices \cite{AvrSad91,Avr95}:

\noindent ({\it i}) continuity
\be\label{CL1}
\psi_{(\alpha\beta_i)}(x_{\alpha\beta_i}=0)=\psi_\alpha
\ \ \mbox{ for }\ \  i=1,\cdots,m_\alpha
\ee
$\{\beta_i \,/\, i=1,\cdots,m_\alpha\}$ is the set of vertices 
neighbours of the vertex $\alpha$~; 
the wave function at the vertex is $\psi_\alpha$.

\noindent ({\it ii}) A second condition sufficient to ensure current 
conservation ({\it i.e.} unitarity of the scattering matrix)
\be\label{CL2}
\sum_\beta a_\ab\,\Dc_{x_{\alpha\beta}}\psi_{(\alpha\beta)}(x_\ab=0)
=\lambda_\alpha\psi_\alpha
\:,\ee
where $\lambda_\alpha$ is a real parameter. Due to the presence of the
connectivity matrix $a_\ab$, the sum runs over all neighbouring vertices 
linked with vertex $\alpha$. 

Note that the conservation of the current alone leads to more general 
boundary conditions and does not require the continuity of the wave function
at the nodes (see \cite{KosSch99,Des01} for example)

The magnetic flux along the bond is denoted by
$\theta_\ab=\int_\alpha^\beta\D x\,A(x)=-\theta_\ba$.

In a scattering situation the graph is connected to the external by leads
plugged on vertices (we designate by ``graph'' the compact part that does
not include the leads).
We call $L$ the number of leads through which some plane wave is injected.
The quantity of interest is the on-shell scattering
matrix $\Sigma$ which is a $L\times L$ matrix that relates the incoming
amplitudes in the $L$ channels to  the outcoming ones. We call
$A^{\rm ext}_\alpha$ (resp. $B^{\rm ext}_\alpha$) the incoming (resp.
outcoming) amplitude on the external lead connected at the vertex $\alpha$
({\it i.e.} $A^{\rm ext}_\alpha$ is the coefficient of a plane wave
$\EXP{-\I kx}$ sent from the lead connected to vertex $\alpha$). By definition:
\be\label{defscatt}
B^{\rm ext} = \Sigma\: A^{\rm ext}
\:.\ee
The purpose of \cite{TexMon01} was to formulate in general terms the
scattering theory for the Schr\"odinger operator, generalizing results
known in the absence of potential $V(x)$ (Laplace operator) 
\cite{AvrSad91,KotSmi99,KotSmi00}.
We have found various expressions of $\Sigma$ for arbitrary graphs and 
related $\Sigma$ to matrices encoding informations about the topology of the 
graph, the potential on the bonds and the way the graph is connected to leads.
We recall here the main results of \cite{TexMon01} that will be necessary
in the following discussion.

\vspace{0.25cm}

\noindent {\bf (A) The arc matrices formulation}.

\noindent We express here $\Sigma$ on the energy shell $E=k^2$ in terms of 
matrices that couple arcs.
The graph is described by $2B$ internal arcs. $L$ external arcs are associated 
to the $L$ leads.
We introduce the matrix $R$ that encodes the information about the potential
on the graph and couples the $2B$ internal arcs of the graph:
\be\label{r1}
R_{ij} = r_i \,\delta_{i,j} 
       + t_{\bar i} \,\delta_{\bar i,j}
\ee
is the matrix element between arcs $i$ and $j$. The potential on each
bond $(i)$ is characterized by reflection and transmission coefficients:
$r_i$, $t_i$ for the injection of the wave in the direction of arc the $i$
and $r_{\bar i}$, $t_{\bar i}$ for the injection in the direction of
the reversed arc $\bar i$.
$R$ is the bond scattering matrix.
If the potential vanishes ($V(x)=0$) we have $r_i=0$ and 
$t_i=\exp(\I kl_i+\I\theta_i)$.

Next we introduce the vertex scattering matrix $Q$ that encodes the
information on the topology of the graph and the way it is connected to
leads:
\bea
Q_{ij} & = & \frac{2}{m_\alpha+\I\lambda_\alpha/k}-1
             \hspace{0.5cm} \mbox{ if } i=j
	     \mbox{ ($i$ issues from the vertex $\alpha$)}  \label{q1}\\
       & = & \frac{2}{m_\alpha+\I\lambda_\alpha/k}
             \hspace{1.15cm} \mbox{ if } i\neq j
	     \ \mbox{ both issuing from the vertex $\alpha$} \label{q2}\\
       & = & 0 \hspace{3cm}\mbox{ otherwise} \label{q3}
\:.\eea
This expression of the vertex scattering matrix is a consequence of the 
conditions (\ref{CL1},\ref{CL2}).
We have also explained in \cite{TexMon01} how this matrix is affected by the
introduction of tunable couplings to the leads.
The $(2B+L)\times(2B+L)$-matrix $Q$ couples the $2B$ internal arcs together
but also the latter to the $L$ external arcs. If we separate $Q$ into 
corresponding blocks:
\be
Q = 
\left(
\begin{array}{c|c}
Q^{\rm int} & \tilde Q^{\rm T} \\ \hline 
\tilde Q   & Q^{\rm ext}
\end{array}
\right)
\ee
then the scattering matrix is:
\be\label{RES1}
\Sigma = Q^{\rm ext} + 
\tilde Q \, ({R^{\dagger} - Q^{\rm int}})^{-1} \, \tilde Q^{\rm T}
\:.\ee
The expression (\ref{RES1}) generalizes the result known in the absence of
potential \cite{KotSmi00}.

\vspace{0.25cm}

\noindent {\bf (B) The vertex matrices formulation}.

\noindent The previous approach is quite natural since we considered 
scattering matrices of the different parts of the system but it has the
disadvantage of dealing with rather big matrices. It is more efficient to
consider matrices that couple vertices.
We define the $L\times V$-matrix $W$ that encodes the information about 
the way the graph is connected to leads:
\be\label{RES4}
W_\ab=w_\alpha\,\delta_\ab
\ee
with $\alpha\in{\cal V}_{\rm ext}$ and $\beta\in{\cal V}$, where
${\cal V}=\{1,\cdots,V\}$ is the set of vertices and 
${\cal V}_{\rm ext}$ the set of vertices connected to leads 
(${\rm Card}({\cal V}_{\rm ext})=L$).
The parameter $w_\alpha\in\RR$ describes the coupling between the graph
and the lead at vertex $\alpha$~; its precise physical meaning is
discussed in \cite{TexMon01}.
In the arc matrices formulation these parameters are introduced in the
matrix $Q$ \cite{TexMon01}.
We just recall that $w_\alpha=1$ corresponds to perfect coupling [the case
considered above in paragraph (A)], whereas $w_\alpha=0$ corresponds to 
disconnect the lead.
The limit $w_\alpha=\pm\infty$ also corresponds to disconnection of the 
lead, however the current is not allowed to flow through the vertex in this 
case and this way to disconnect the lead is equivalent to impose a 
Dirichlet boundary at the vertex ($\lambda_\alpha=\infty$).

We also introduce the matrix $M$ that contains all the information on the 
isolated graph (potential on the bond and topology):
\bea\label{RES3}
M_\ab &=&
\delta_\ab\left(\I\frac{\lambda_\alpha}{k}
+ \sum_\mu a_{\alpha\mu}
\frac{(1-r_{\alpha\mu})(1+r_{\mu\alpha})+t_{\alpha\mu}\,t_{\mu\alpha}}
     {(1+r_{\alpha\mu})(1+r_{\mu\alpha})-t_{\alpha\mu}\,t_{\mu\alpha}}
\right)
\nonumber \\ && \hspace{1cm}
- a_\ab\frac{2\,t_\ab}{(1+r_\ab)(1+r_\ba)-t_\ab\,t_\ba}
\:.\eea
Then, the scattering matrix reads:
\be\label{RES2}
\Sigma = -1 + 2 \, W \left( M + W^{\rm T}W\right)^{-1} W^{\rm T}
\:.\ee
These equations generalize the result known in the absence of the
potential \cite{AvrSad91,KotSmi99}. In this latter case we recover from 
(\ref{RES3}) the well-known matrix:
\be\label{fmM}
M_\ab = \I\,\delta_\ab\sum_\mu a_{\alpha\mu} \cotg kl_{\alpha\mu}
-a_\ab\frac{\I\,\EXP{\I\theta_\ab}}{\sin kl_\ab}
\:.\ee

Some examples of application of these formulae are given in \cite{TexMon01}.

\vspace{0.25cm}

We describe the organization of the paper.
Since there has been recently some confusion in the literature about
the content of the Friedel Sum rule, we think it is useful to spend some
time by reviewing some aspects around this relation, which will be also
necessary for the following.
In section \ref{sec:WSR} we generalize the Smith formula for 
graphs. Then we provide in section \ref{sec:VoFSR} several examples of 
violation of the Friedel sum rule and explain the origin of this failure.


\section{The Friedel Sum Rule \label{sec:FSR}}

To be precise we consider the scattering theory for the Schr\"odinger 
equation on a three dimensional Euclidean manifold and restrict ourselves
to the case of a rotational invariant potential supposed to be concentrated 
in a sphere of radius $R$.
A basis of eigenstates is given by the partial waves 
$\psi_{l}(r)\,Y_l^m(\theta,\varphi)$ [where $Y_l^m(\theta,\varphi)$ are
the spherical harmonics] whose radial parts involve the phase shifts 
$\eta_l(E)$:
$\psi_{l}(r)=\frac{1}{\sqrt{\pi k}}\,\frac{1}{r}\sin(kr -l\pi/2 +\eta_l)$
for $r\geq R$. The energy of this eigenstate is $E=k^2$.
The Krein-Friedel relation relates the variation of the density of states 
(DoS) to scattering properties. We introduce the local density of states
(LDoS) $\rho(\vec r;E)=\bra{\vec r}\delta(E-H)\ket{\vec r}$. We denote
$\rho_0(\vec r;E)$ the LDoS in the absence of the potential.
The relation reads:
\be\label{RBU37}       
\int\D\vec r\,\left[ \rho(\vec r;E)-\rho_0(\vec r;E) \right] 
= \frac1\pi\sum_{l=0}^\infty(2l+1)\frac{\D\eta_l}{\D E}
\:.\ee
Since the integral in the l.h.s. runs over the whole space, the total density 
of states (DoS) is diverging like the volume of integration; however the 
difference of the l.h.s. is a finite quantity. The demonstration of 
(\ref{RBU37}) in the one-dimensional case is recalled in appendix 
\ref{app:FSR1D}.

Using the fact that $\EXP{2\I\eta_l}$ are the eigenvalues of the on-shell
scattering matrix $\tilde\Sigma$ we can write\footnote{
  The scattering matrix $\tilde\Sigma$ entering into (\ref{RBU}) is slightly
  different from the scattering matrix $\Sigma$ we have introduced, both being
  related through a simple transformation.
  The phase shifts $\eta_l(E)$ (phases of the eigenvalues $\EXP{2\I\eta_l}$ of
  $\tilde\Sigma$) encode the effect of the scattering potential compared with
  the free case: in the absence of the potential, the phase shifts $\eta_l$ 
  vanish.
  On the other hand the phases $\delta_l(E)$ of the eigenvalues
  $\EXP{\I\delta_l}$ of $\Sigma$ are measured from the edge of the scattering
  region:
  $\psi_{l}(r)
  =\frac{1}{\sqrt{\pi k}}\,\frac{1}{r}\sin(k(r-R) -l\pi/2 +\delta_l/2)$
  for $r\geq R$.\\
  Therefore we have: $\delta_l=2\eta_l+2kR$. The relation between $\Sigma$
  and $\tilde\Sigma$ is also explained in detail in the one-dimensional case
  in appendix \ref{app:FSR1D}.}:
\be\label{RBU}
\int\D\vec r\,\left[ \rho(\vec r;E)-\rho_0(\vec r;E) \right] 
=\frac{1}{2\I\pi} \frac{\D}{\D E} {\rm Tr}\{\ln\tilde\Sigma(E)\}
=\frac{1}{2\I\pi} \frac{\D}{\D E} \ln\det\tilde\Sigma(E)
\:,\ee
where the trace is computed on the energy shell $E$ over channel indices.

It is convenient to introduce the Friedel phase defined as: 
$\delta^f(E)=-\I\ln\det\Sigma(E)$ with the additional constraint to be a 
continuous function of the energy. 
It is the sum of the cumulative phases of 
the eigenvalues $\EXP{\I\delta_a}$ of the scattering matrix $\Sigma$: 
$\delta^f(E)=\sum_{a}\delta_a(E)$.
The Friedel phase counts the number of resonance peaks: if they are
sufficiently narrow to be well separated, 
in the neighbourhood of a resonance, the determinant behaves like
\be
\EXP{\I\delta^f(E)}=\det\Sigma(E) 
{{\raisebox{-.3cm}{$\textstyle\propto$}} \atop {\scriptstyle{E\sim E_n}}}
\left(\frac{E-E_n - \I\Delta_n}{E-E_n + \I\Delta_n}\right)^{d_n}
\:,\ee
up to a constant phase.
$E_n$ is the position of the resonance, $\Delta_n$ its width and
$d_n$ the degeneracy of the state. This expression shows that the phase
$\delta^f(E)$ makes a jump of $2\pi d_n$ when $E$ crosses the resonance.

The relation (\ref{RBU37}) was derived long ago by Beth and Uhlenbeck 
\cite{BetUhl37} in the context of the study of a gas of interacting 
particles, where it is involved in the second virial coefficient 
(related to the two-body problem). 
The generalization for a systematic expansion of the grand potential was 
provided in \cite{DasMaBer69}. The demonstration of the Krein-Friedel relation 
\cite{Fri52,Kre53,LanAmb61,Bus62}, also called Friedel sum rule, 
is given in standard textbooks for rotational invariant potentials 
\cite{Hua63,LanLif66e}.
It is also worth mentionning the existence of a vast literature in 
mathematical physics dealing with the scattering theory. Many references 
can be found in \cite{Yaf92} which devotes its last chapter to the study
of the Krein spectral shift function (the Friedel phase) and trace formula.
The matrix $-\I\Sigma^\dagger\frac{\D\Sigma}{\D E}$ whose trace is 
computed in (\ref{RBU}) is the matrix of Wigner time delays (note also the 
existence of a classical formulation of the second virial coefficient 
in terms of classical time delays in \cite{Ma85}).
It is worth mentionning that (\ref{RBU}) is exact which is the beauty of
this relation (its validity is not restricted to a high energy regime for 
example). 
Integrated over the interval of energy below the Fermi energy, 
(\ref{RBU37},\ref{RBU}) give the accumulation of charge due to the presence 
of the potential, to use the language of \cite{Fri52}.

Instead of considering the variation of the DoS of the whole space, it is 
also possible to study the LDoS integrated over the interacting region only. 
This quantity is also related to scattering properties through the 
Smith formula \cite{Smi60} which defines the time delay. For a rotational 
invariant potential the relation reads:
\be\label{Smith60}
2\pi\int_{0}^{R}\D r\,r^2 |\psi_l(r)|^2 =  
2\frac{\D\eta_l}{\D E} 
+\frac{R}{k} - \frac{1}{2E}\sin(2kR+2\eta_l-l\pi)
=\frac{\D\delta_l}{\D E} - \frac{1}{2E}\sin(\delta_l-l\pi)
\:.\ee
Note that this relation was also derived in \cite{Fri52} as an intermediate 
result for the demonstration of (\ref{RBU37}).
With a summation over the angular quantum numbers\footnote{
  The choice of normalization for the stationary scattering states 
  $\psi_{E,l,m}=\psi_l(r)Y_l^m(\theta,\varphi)$ introduced above corresponds
  to associate to those states a measure $\D E$ [it implies for example that
  $\int\D\vec r\,\psi_{E,l,m}^*\psi_{E',l',m'}
   =\delta_{l,l'}\delta_{m,m'}\delta(E-E')$].}, 
we get the LDoS integrated over the sphere:
\be\label{Smith60bis}
\int_{r<R}\D\vec r\,\rho(\vec r;E)
=\sum_{l=0}^\infty(2l+1)\int_{0}^{R}\D r\,r^2 |\psi_l(r)|^2
=\frac{1}{2\pi}\sum_{l=0}^\infty(2l+1)
\left(\frac{\D\delta_l}{\D E} - \frac{1}{2E}\sin(\delta_l-l\pi)\right)
\:.\ee
If the coupling between the scattering region (sphere of radius $R$) and 
the external is adjustable, this quantity corresponds to the DoS of the 
scattering region when it is isolated.
If we are interested in the Weyl contribution of the DoS of the scattering 
region we can forget the second term of (\ref{Smith60bis}) and consider only 
the Weyl term of the Friedel phase.

Due to the central position of the scattering approach in mesoscopic physics,
the FSR plays an important role in the study of many physical quantities:
for example the FSR allows to relate the persistent current to 
scattering properties \cite{AkkAueAvrSha91} and is also involved in 
electrochemical capacitance \cite{But93,GopMelBut96}. 
A local formulation was also developed in \cite{But93,GasChrBut96} to 
relate the local density of states to scattering properties (a general 
discussion of the role of the local FSR is provided in \cite{But02}).
Since graphs are widely used to model mesoscopic networks they have 
been considered to apply concepts involving the FSR, like in
\cite{AkkAueAvrSha91} for the persistent current in a loop connected to
one lead, or in  the recent work \cite{TanBut99} in which graphs provided
examples to illustrate a general discussion about some subbtle point related
to phases.

Recently there has been some confusion about the FSR in \cite{DeoBanDas01}.
Starting from a misinterpretation of the FSR, these authors claimed
that the relation does not hold in the one-dimensional case if the 
potential is made of two $\delta$ peaks, which is not true. We repeat that
the general demonstration of (\ref{RBU}) in \cite{DasMaBer69} covers the 
one-dimensional situation.
The one-dimensional case is reviewed in detail in appendix
\ref{app:FSR1D} where we consider as an example the case of one $\delta$
peak (it is not difficult to check that the FSR works perfectly well, as 
it should, for two $\delta$ peaks, a little exercice following the same 
lines).

The FSR has been proven in arbitrary dimension however it has 
not been demonstrated for graphs which are intermediate objects between one
dimensional and higher dimensional systems. Then it is important to clarify 
some points in this context.
The FSR (\ref{RBU37},\ref{RBU}) counts the variation of the DoS due to a 
scattering potential. The Smith relation (\ref{Smith60bis}) measures the 
LDoS integrated in the scattering region. Both are based on the idea to count 
the number of states of the scattering region by counting the resonance 
peaks of the phase shifts derivatives.
We will show that this procedure is not always applicable for graphs:
for example in the case of the complete graph that will be studied in
detail below, some states of the isolated graph are not manifesting by a 
resonance peak or give rise to a resonance peak that does not carry the 
correct spectral weight (the degeneracy of the level). 
To study this problem it will be sufficient to consider the Weyl part of 
the Friedel phase to notice some discrepancy with the Weyl part of the 
DoS of the graph.
Before following this program and to settle the discussion on more precise
grounds, we will generalize the Smith formula (\ref{Smith60}) to the case of
graphs, despite it does not always concern the DoS as we will see.


\section{Generalization of the Smith relation\label{sec:WSR}}

The Smith relation was derived for a one-dimensional system with one 
scattering channel \cite{Smi60} (or rotational invariant potentials 
in 3 dimensions) and involves the Wigner time delay \cite{Wig55}.
We generalize this relation to the case of a perfectly connected graph 
($w_\alpha=1$).
The starting point is to introduce
\be\label{defom}
\Omega = \left(\Dc_x\psi\right)^*\frac{\D\psi}{\D E}
-\psi^*\left(\Dc_x\frac{\D\psi}{\D E}\right)
\:\ee
which satisfies the following relation:
\be\label{pws1}
\frac{\D}{\D x} \Omega(x) = |\psi(x)|^2
\ee
for any solution $\psi$ of the Schr\"odinger equation
$(-\Dc^2_x+V(x))\psi(x)=E\psi(x)$~; we recall that $\Dc_x=\D_x-\I A(x)$ is 
the covariant derivative. Applied to a graph, the relation (\ref{pws1}) 
should be written for the $B$ components of the wave function on the bonds 
(and also on the $L$ leads).

We first derive two properties involving $\Omega_\ab(x_\ab)$, the quantity 
(\ref{defom}) related to the component of the wave function 
$\psi_{(\ab)}(x_\ab)$ (we will denote $\Omega_{{\rm lead}\:\mu}(x)$ the one 
associated with the component $\psi_{{\rm lead}\:\mu}(x)$ on the lead 
attached to the vertex $\mu$). 
Obviously, we have $\Omega_\ab(x_\ab)=-\Omega_\ba(x_\ba)$. 
Due to the conservation of the current, ensured by (\ref{CL2}), the sum of 
all $\Omega_\mb$'s associated to the arcs issuing from the vertex $\mu$ 
and computed at the position of the vertex ($x_\mb=0$), is zero:
\be\label{pws2}
\sum_\beta a_\mb\,\Omega_\mb(\mu) 
+ (W^{\rm T}W)_{\mu\mu}\, \Omega_{{\rm lead}\:\mu}(\mu)= 0
\:.\ee
We have used the obvious notation: 
$\Omega_\mb(\mu)\equiv\Omega_\mb(x_\mb=0)$.
The second term is the contribution of a lead, if one is plugged at 
vertex $\mu$ [due to the definition of $W$, we recall that 
$(W^{\rm T}W)_{\mu\mu}=1$ if a lead issues from $\mu$ and $0$ otherwise].
The second useful property is obtained by integration of (\ref{pws1}) on the
bond $(\mb)$:
\be\label{pws3}
\int_0^{l_\mb}\D x\,|\psi_{(\mb)}(x)|^2 
= -\Omega_\mb(\mu) -\Omega_\bm(\beta)
\:.\ee

We now consider the stationary scattering state $\psi^{(\alpha)}(x)$ of 
energy $E=k^2$, associated to the injection of a plane wave from the lead
connected at vertex $\alpha$. The construction of these eigenstates is
briefly recalled in appendix \ref{app:SSS} (see \cite{TexMon01}). From the 
expression (\ref{wfl}) of the wave function on the lead we see that
\be\label{pws4}
\Omega_{{\rm lead}\:\mu}^{(\alpha)}(\mu) =
-2\I k\, \Sigma_{\mu\alpha}^* \frac{\D \Sigma_{\mu\alpha}}{\D E}
-\frac{\I}{2 k} \left(\delta_{\mu\alpha}+\Sigma_{\mu\alpha}^*\right)
\left(-\delta_{\mu\alpha}+\Sigma_{\mu\alpha}\right)
\:.\ee
We now compute the integral of $|\psi^{(\alpha)}(x)|^2$ on the graph (the 
``graph'' refers to internal bonds):
\be\label{pws5}
\int_{\rm Graph} \D x\,|\psi^{(\alpha)}(x)|^2
=\sum_{(\mb)}\int_0^{l_\mb}\D x\,|\psi_{(\mb)}^{(\alpha)}(x)|^2
=-\sum_{{\rm arc}\:\mu\nu} \Omega_{\mu\nu}^{(\alpha)}(\mu)
\:,\ee
where we have used (\ref{pws3}).
The summation $\sum_{(\mb)}$ is over the $B$ bonds of the graph whereas the 
last summation runs over the $2B$ internal arcs. We see that the contributions 
from the arcs issuing from an internal vertex vanish due to (\ref{pws2}). 
The contributions of the internal arcs issuing from connected vertices 
can be replaced by the contributions of external leads due to (\ref{pws2}).
Therefore we obtain:
\be\label{pws6}
\int_{\rm Graph} \D x\,|\psi^{(\alpha)}(x)|^2
=\sum_\mu \Omega_{{\rm lead}\:\mu}^{(\alpha)}(\mu)
=-2\I k\,\left(\Sigma^\dagger\frac{\D\Sigma}{\D E}\right)_{\alpha\alpha}
       -\frac{\I}{2k}(\Sigma_{\alpha\alpha}-\Sigma_{\alpha\alpha}^*)
\ee
where the sum over $\mu$ is obviously over the $L$ connected vertices.

In order to associate a measure $\D E$ to the stationary scattering states,
we change the normalization. The scattering states (\ref{wfl},\ref{wfb}) are 
related to the new ones by:
$\tilde\psi_E^{(\alpha)}(x) = \frac{1}{\sqrt{4\pi k}}\psi^{(\alpha)}(x)$.
If we sum the contributions (\ref{pws6}) of the $L$ stationary scattering 
states of energy $E$, we obtain:
\begin{center}
\begin{tabular}{|p{14cm}|}
\hline
\be\label{SmiG}
\sum_\alpha \int_{\rm Graph} \D x\, |\tilde\psi_E^{(\alpha)}(x)|^2
= \frac{1}{2\I\pi}
\left(
  \tr{\Sigma^\dagger\frac{\D\Sigma}{\D E}} 
+ \frac{1}{4E}\tr{\Sigma-\Sigma^\dagger}\right)
\ee
\\
\hline
\end{tabular}
\end{center}
which generalizes the Smith relation (\ref{Smith60}) to the case of 
graphs. The term $-\I\tr{\Sigma^\dagger\frac{\D\Sigma}{\D E}}$
is the time delay.

To compute the Friedel phase of a graph appearing in the above relation,
it is useful to note that\footnote{
  The proof is easily achieved by considering the graph ${\cal G}'$ related
  to the original graph ${\cal G}$ by attaching to each of the $V-L$ internal
  vertices of ${\cal G}$ (labelled for convenience with prime indices:
  $\alpha'$,\ldots ) a lead with tunable coupling. If these couplings are 
  swichted off ($w_{\alpha'}\to0$), the $V\times V$ scattering matrix 
  $\Sigma'$ of ${\cal G}'$ is
  block diagonal with a $L\times L$ block being the scattering matrix $\Sigma$
  of ${\cal G}$, the other $(V-L)\times(V-L)$ block corresponding to the
  additional leads being $-1$.
  Let us now compute $\det\Sigma'$: for finite couplings $w_{\alpha'}$, the
  matrix $W'$ describing the coupling of ${\cal G}'$ to the $V$ leads is
  square and possesses an inverse. It follows from (\ref{RES2}) that
  $\Sigma'=W'^{-1}(W'^2-M)(W'^2+M)^{-1}W'$. Then
  $\det\Sigma' = \frac{\det(W'^2-M)}{\det(W'^2+M)}$.
  If the couplings to the additional leads now vanish, $w_{\alpha'}\to0$, 
  we have $\det\Sigma'=(-1)^{V-L}\det\Sigma$ and $W'^2=W^{\rm T}W$. 
  {\sc Qed}.}
\be\label{detS}
\det\Sigma = (-1)^{V-L} \frac{\det(W^{\rm T}W-M)}{\det(W^{\rm T}W+M)}
\:.\ee

For one channel (one lead) we have $\Sigma=\EXP{\I\delta}$, therefore we get
\be
\int_{\rm Graph} \D x\, |\tilde\psi_E(x)|^2
= \frac{1}{2\pi}
\left(
  \frac{\D\delta}{\D E} 
+ \frac{1}{2E}\sin\delta\right)
\:,\ee
which would be the relation (\ref{Smith60}) obtained by Smith if the graph
would reduce to a line (one-dimensional case with one channel). The
different sign of the second term is only a matter of definition of
the phase shift $\delta$, in the one-dimensional case, and $\delta_0$, in
the $l=0$ channel of the three-dimensional case, which differ by $\pi$.

\mathversion{bold}
\subsubsection*{Case $w_\alpha\neq1$}
\mathversion{normal}

When we introduce arbitrary couplings between the leads and the graph, the 
application
of formula (\ref{SmiG}) means that we are also taking into account integral
over the bonds on which are the barriers characterized by $w_\alpha$'s
(see \cite{TexMon01} where the introduction of these parameters is explained
in detail).


\section{Violation of the Friedel sum rule for certain graphs
         \label{sec:VoFSR}}

The idea of the FSR is to count the states in the scattering region by
studying the Friedel phase.
We have seen that the Smith formula (\ref{Smith60}) relates the LDoS
integrated over the scattering region to the Friedel phase and we have found
its generalization (\ref{SmiG}) for graphs.
We call ${\cal N}(E)=\int_{-\infty}^E\D E'\int_{\rm Graph}\D x\,\rho(x;E')$
the integrated density of states (IDoS) of the graph.
If we are not interested in the details of the spectrum but only in the
Weyl term of the IDoS of the scattering region, and if we believe the FSR, 
the relation (\ref{Smith60bis}) shows that 
${\cal N}_{\rm Weyl}(E)\simeq\frac{1}{2\pi}\delta^f(E)$ up to some
oscillatory part.
As a matter of fact this is not always true for graphs and we will now give
several examples where ${\cal N}_{\rm Weyl}(E)$ is not given by the dominant 
contribution of $\delta^f(E)$.

All the examples we are going to consider are free graphs, with $V(x)=0$, but 
the ideas that will come out are not specific to free graphs. 
We recall that in the absence of a potential we have\footnote{
The Weyl term appears in the trace formula originally derived by Roth 
\cite{Rot83} (see also \cite {KotSmi99,AkkComDesMonTex00}).
}:
${\cal N}_{\rm Weyl}(E)=\frac{{\cal L}k}{\pi}$, where
${\cal L}=\frac12\sum_{\alpha,\beta}a_\ab l_\ab$ is the total length of the
graph.
As we will see, one of the reasons of the violation of the FSR is the
occurence of
degeneracies in the spectrum. Graphs with symmetries present a lot of
degeneracies. This is why it is interesting to start by studying the
complete graph $K_V$, which is the most symmetric simply connected graph
with $V$ vertices.


\mathversion{bold}
\subsection{The complete graph $K_V$ connected to one lead.} 
\mathversion{normal}

The graph $K_V$ is made of $V$ vertices, each being connected to the other 
ones by bonds of same length $\ell$. The matrix $M$ takes a simple form
(see equation (\ref{MKn}) in appendix \ref{app:GC}).

\begin{figure}[!h]
\begin{center}
\includegraphics[scale=0.6]{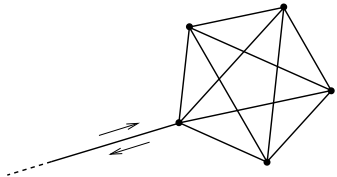}
\end{center}
\caption{Complete graph $K_5$ connected to one lead.}\label{fig:k5s}
\end{figure}

The graph is connected to one lead (figure \ref{fig:k5s}) and the scattering 
matrix is characterized by a unique phase: $\Sigma=\EXP{\I\delta^f}$.
Using (\ref{RES2}) we recover after a little bit
of algebra the expression \cite[formula (119)]{AkkComDesMonTex00}:
\be\label{pscg}
\cotg(\delta^f/2)
=\cos\varphi\frac{\cos k\ell +\cos\varphi -1}{\cos k\ell +\cos\varphi }
\cotg(k\ell/2)
\:,\ee
where $k=\sqrt{E}$ and $\cos\varphi=\frac{1}{V-1}$. 
This expression shows that $\delta^f(k^2)=3k\ell+({\rm fluct.})$, where 
$({\rm fluct.})$ represents a fluctuating term of order $\pi$, whereas the
Weyl part of the IDoS is 
${\cal N}_{\rm Weyl}(E)=\frac{B\ell k}{\pi}=\frac{V(V-1)k\ell}{2\pi}$,
which clearly shows the discrepancy between ${\cal N}(E)$ and 
$\frac{\delta^f(E)}{2\pi}$.

A more detailed analysis of the position of the resonance peaks of
$\frac{\D\delta^f}{\D E}$ shows that the Friedel phase does not measure 
the degeneracies of the energies of the isolated graph (see appendix 
\ref{app:GC} where the spectrum of $K_V$ is recalled) and moreover even 
misses some energies: there is no resonance peak at $k_{2+4n}$.

To understand in more general terms the origin of the failure of the FSR
when only one lead is plugged on the graph, we consider the simple case
of a graph with no potential ($V(x)=0$ and $\lambda_\alpha=0$) 
connected to only one external lead. The formula 
\cite{AkkComDesMonTex00}
\be\label{rpd}
\cotg(\delta^f(E)/2)
=-\sqrt{E}\frac{S_{\rm Dir.}(-E-\I0^{+})}{S_{\rm Neu.}(-E-\I0^{+})}
\ee
relates the phase shift $\delta^f(E)$ to the ratio of two spectral
determinants. 
On the one hand $S_{\rm Dir.}(\gamma)$ is the spectral determinant
$\det(-\Dc^2_x+\gamma)$ calculated with a Dirichlet boundary condition
($\lambda_{\alpha_0}=\infty$) at the vertex $\alpha_0$ where the lead 
is plugged in, and Neumann boundary conditions at all other vertices 
($\lambda_\alpha=0$). 
On the other hand $S_{\rm Neu.}(\gamma)$ is calculated with Neumann boundary 
conditions at all vertices.
The sum rule means that each state in the isolated graph is associated to a 
jump of $2\pi$ of the phase $\delta^f$. Due to
(\ref{rpd}) we see that a jump of $2\pi$ occurs when the expression
(\ref{rpd}) diverges. Then we identify two reasons why the FSR fails:
({\it i}) if the spectrum of the graph is degenerate and ({\it ii}) if
$S_{\rm Dir.}(\gamma)$ vanishes for the same energy as $S_{\rm Neu.}(\gamma)$, 
then (\ref{rpd}) diverges a number of times which is not related to the
number of states in the graph.


\mathversion{bold}
\subsection{The complete graph $K_V$ connected to $V$ leads.}
\mathversion{normal}

\begin{figure}[!h]
\begin{center}
\includegraphics[scale=0.8]{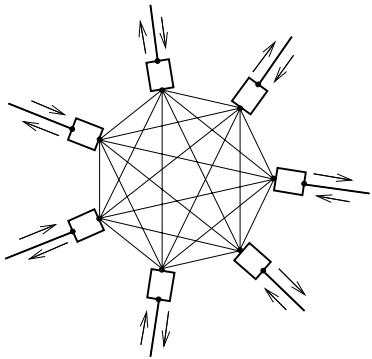}
\end{center}
\caption{Complete graph $K_7$ connected to 7 leads. The small boxes
represent the couplings characterized by the parameter $w$ \cite{TexMon01}.}
\label{fig:k7s}
\end{figure}

To convince ourself that the breakdown of the Krein-Friedel relation is 
not specific to graphs connected to one lead only, we consider now the case
where $K_V$ is attached to $V$ leads connected to each vertices 
(figure \ref{fig:k7s}).

If all the vertices of the graph are connected to leads ($L=V$), the matrix 
$W$ is square. The determinant (\ref{detS}) is:
$\EXP{\I\delta^f(E)}=\frac{\det(W^2-M)}{\det(W^2+M)}$. For simplicity we
consider the case of equal couplings: $w_1=w_2=\cdots=w_V=w$.
Using (\ref{MKn}) we get:
\be\label{FFKn}
\EXP{\I\delta^f(k^2)}=(-1)^V
\frac{\sin(k\ell/2)-\I w^2\cos\varphi\cos(k\ell/2)}
     {\sin(k\ell/2)+\I w^2\cos\varphi\cos(k\ell/2)}
\left(
\frac{\cos k\ell+\cos\varphi+\I w^2\cos\varphi\sin k\ell}
     {\cos k\ell+\cos\varphi-\I w^2\cos\varphi\sin k\ell}
\right)^{V-1}
\:.\ee
This expression shows that $\delta^f(k^2)=(2V-1)k\ell +({\rm fluct.})$
which disagrees once again with 
${\cal N}_{\rm Weyl}(E)=\frac{V(V-1)k\ell}{2\pi}$.

It is interesting to provide a more detailed analysis by studying the
behaviour of the Friedel phase in the neighbourhood of the energies of the
graph (the spectrum is recalled in appendix \ref{app:GC}). We consider the
limit $w\to0$ for which the resonance profile of 
$\frac{\D\delta^f}{\D E}$ emerges clearly.

\noindent$\bullet$
Near the first energy level (for $k\sim k_1$) we see from (\ref{FFKn}) that
$\EXP{\I\delta^f}\propto
\left(\frac{k-k_1-\I\Delta k_1}{k-k_1+\I\Delta k_1}\right)^{V-1}$
with $\Delta k_1\ell=w^2/(V-1)$. The exponent is the degeneracy of the level
which means that the resonance peak of $\frac{\D\delta^f}{\D E}$ has the
correct spectral weight and counts correctly the $V-1$ states.

\noindent$\bullet$
In the neighbourhood of the second  energy level (for $k\sim k_2$)
$\frac{\D\delta^f}{\D E}$ is flat: $\delta^f$ is not sensitive to the
presence of states at this energy.

\noindent$\bullet$
The situation at $k\sim k_3$ is the same as the one at $k\sim k_1$ 
(with $\Delta k_3=\Delta k_1$)

\noindent$\bullet$
At $k\sim k_4$ we have 
$\EXP{\I\delta^f}\propto
\frac{k-k_4-\I\Delta k_4}{k-k_4+\I\Delta k_4}$ 
with $\Delta k_4=2\Delta k_1$: the Friedel phase misses the degeneracy.

One may now ask the question why the Friedel phase misses some states
sometimes and sometimes not~? To answer this question we can study the
structure of the wave functions of the isolated graph (see appendix
\ref{app:GC}). Whereas the wave function is finite at the nodes at energies
$k_{1+2m}$ where the Friedel phase is sensitive to the degeneracy,
all the $V(V-3)/2$ degenerate wave functions vanish at all the nodes at
energies $k_{2+4m}$ as well as at energies $k_{4+4m}$, which means that the
wave sent from the lead does not enter the graph at those energies.


\subsection{The ring connected with two leads.}

\begin{figure}[!h]
\begin{center}
\includegraphics[scale=0.9]{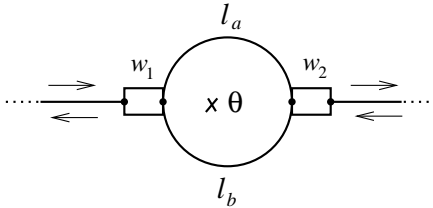}
\end{center}
\caption{The ring connected to two leads and threatened by a flux $\theta$.
The two arms have lengths $l_a$ and $l_b$. The parameters $w_1$ and $w_2$
allow to tune the coupling of the ring.}
\label{fig:ring5}
\end{figure}

To understand better the remark that closed the previous subsection we
consider next a simpler case: a ring connected to two leads (figure
\ref{fig:ring5}). The arms of the ring are of length $l_a$ and $l_b$ with
$l=l_a+l_b$.

Let us first recall the spectrum of the isolated ring of perimeter $l$
threatened by a flux $\theta$: the energies are
$E_m(\theta)=\left(\frac{2\pi}{l}\right)^2\left(m-\frac{\theta}{2\pi}\right)^2$
associated to wave functions
$\varphi_m(x)=\frac{1}{\sqrt{l}}\EXP{2\I\pi mx/l}$ for $m\in\ZZ$.
If $\theta=0$ the states $\varphi_m$ and $\varphi_{-m}$ are degenerate.
Therefore we can introduce in this case a different basis:
a symmetric function 
$
\varphi_n^+=\frac{\varphi_n+\varphi_{-n}}{\sqrt2}
=\sqrt{{2}/{l}}\cos(2\pi nx/l)
$ 
with $n\in\NN$, and an antisymmetric one:
$
\varphi_n^-=\frac{\varphi_n-\varphi_{-n}}{\I\sqrt2}
=\sqrt{{2}/{l}}\sin(2\pi nx/l)
$ 
with $n\in\NN^*$.

\noindent
We now consider the graph when it is coupled to two leads with coupling
parameters $w_1$ and $w_2$ (see figure \ref{fig:ring5}).
The Friedel phase $\EXP{\I\delta^f}=\frac{\det(W^2-M)}{\det(W^2+M)}$ is:
\be
\EXP{\I\delta^f} = 
\frac{2(\cos\theta-\cos kl)+w_1^2w_2^2\sin kl_a\sin kl_b
      -\I(w_1^2+w_2^2)\sin kl}
     {2(\cos\theta-\cos kl)+w_1^2w_2^2\sin kl_a\sin kl_b
      +\I(w_1^2+w_2^2)\sin kl}
\ee
($\det(W^2+M)$ was calculated in \cite{TexMon01}).
If $\theta\neq0$ the spectrum is non degenerate and $\delta^f$ counts
correctly the states. Now we focus on the degenerate case $\theta=0$ for
which we have:
\be\label{tdfr}
\tan(\delta^f/2) = 
-\frac{(w_1^2+w_2^2)\sin kl}{4\sin^2(kl/2)+w_1^2w_2^2\sin kl_a\sin kl_b}
\:.\ee
Each interval of width $\Delta k=2\pi/l$ contains 2 states of the ring. Let 
us now examine under what condition (\ref{tdfr}) counts correctly these states.
We can identify the position of the resonances with the value of $k$ for
which the denominator of (\ref{tdfr}) vanishes\footnote{
  The denominator is of the form:
  \be\label{fa}
  f_a(x)=\sin^2({x}/{2}) + b\sin(ax)\sin((1-a)x)
  =(1+b)\sin^2(x/2) - b \sin^2((1/2-a)x)
  \:,\ee
  with $a\in]0;1/2[$ and $b\in]0;+\infty[$. 
  We are interested in the number of zeros of $f_a(x)$ in the interval
  $[2m\pi;2(m+1)\pi[$.
  We distinguish two cases:

  \noindent$\bullet$
  $a$ is not a rational number ($a\notin\QQ$). Since
  $f_a(2m\pi)<0$ $\forall m\in\NN$ and the amplitude of the first positive term 
  in the r.h.s of (\ref{fa}) is larger than the second, it follows that 
  $f_a(x)=0$ has exactly two solutions in $[2m\pi;2(m+1)\pi[$.

  \noindent$\bullet$
  $a\in\QQ$~: we write $a=\frac{p}{2q}$ where $(p,q)\in\NN^2$ with $p<q$.
  We have $f_a(2m\pi)=-b\sin^2((q-p)m\pi/q)\leq0$.\\
  If $(q-p)(m+1)$ is not an integer multiple of $q$ the interval 
  $[2m\pi;2(m+1)\pi[$ contains two solutions of $f_a(x)=0$.\\
  If $(q-p)(m+1)=rq$ with $r\in\NN$ the
  interval contains only one solution of $f_a(x)=0$.}.
Then we distinguish two different cases:

\noindent$\bullet$ $l_a/l$ is an irrational number.
Then the denominator of (\ref{tdfr}) vanishes twice per interval
$k\in[2m\pi/l;2(m+1)\pi/l[$, which means that $\delta^f$ counts the correct
number of states.

\noindent$\bullet$ $l_a/l$ is a rational number: $l_a/l=\frac{p}{2q}$
with $(p,q)\in\NN^2$. If $(q-p)(m+1)$ is an integer multiple of $q$, 
the denominator vanishes only once in $[2m\pi/l;2(m+1)\pi/l[$.
The intervals for which $(q-p)(m+1)$ is an integer multiple of $q$ are 
those in which one of the two degenerate wave functions $\varphi^+_{m+1}$ 
and $\varphi^-_{m+1}$ vanishes on the vertices where the lead are plugged in.


\subsection{The ring connected with one lead.
            Why the l.h.s of (\ref{SmiG}) can not always be identify with 
	    the DoS~?}

We have given a general argument to explain how the degeneracies of the 
spectrum lead to a failure of the FSR, however it is surprising that 
the quantity in the left hand side of (\ref{SmiG}) can not always be 
identified with the DoS of the graph since the sum runs over the complete
set of $L$ stationary scattering states of energy $E$. 
This point needs a clarification that we will give now by studying again 
the case of the ring.

\begin{figure}[!h]
\begin{center}
\includegraphics[scale=0.9]{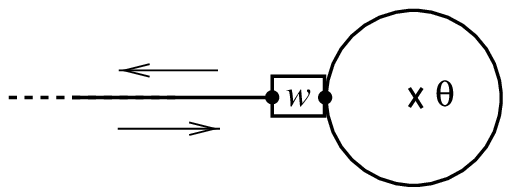}
\end{center}
\caption{}
\label{fig:ring3}
\end{figure}

We consider the ring of figure \ref{fig:ring3} coupled to a lead
and construct the stationary scattering states 
(\ref{wfl},\ref{wfb}). The wave function on the lead is
\be
\tilde \psi^{\rm lead}_E(x) = \frac{1}{\sqrt{4\pi k}}
\left(\EXP{-\I kx}+\EXP{\I kx+\I\delta}\right)
\:,\ee
where the phase shift $\delta$ is \cite{TexMon01}
\be
\cotg(\delta/2) = \frac{w^2\sin kl}{2(\cos\theta - \cos kl)}
\:.\ee
On the ring
(arc $a$) the wave function reads:
\be\label{wfr1}
\psi_{(a)}(x) = \psi_1 \frac{\EXP{\I \theta x/l}}{\sin kl}
\left(\sin k(l-x) + \EXP{-\I\theta}\sin kx\right)
\ee
where 
\be\label{wfr2}
\psi_1 = \frac{1}{\sqrt{\pi k}}\,
\frac{w \sin kl}{w^2\sin kl+2\I(\cos kl-\cos\theta)}
=\frac{1}{w}\tilde \psi^{\rm lead}_E(0)
\ee
is the wave function at the node.

\noindent$\bullet$ \mathversion{bold}$\theta\neq0$.\mathversion{normal}
We study the limit $w\to0$ of small coupling for which we expect to recover
some features of the isolated ring. In this case $|\psi_1|$
presents sharp peaks at the positions of the energies of the isolated ring
(given by $k_n^\pm l=\pm\theta+2n\pi$). These resonance
contributions will eventually give the main contributions to (\ref{SmiG}). 
Let us express the wave function in the ring for $k$ in the neighbourhood of a 
resonance. Expressions (\ref{wfr1},\ref{wfr2}) give:
\be
\psi_{(a)}(x) \APPROX{k\sim k_n^\pm} \frac{1}{\sqrt{\pi k}}\, 
\frac{\I w/2l}{k-k_n^\pm + \I w^2/2l}\, \EXP{\mp2\I\pi nx/l}
\:.\ee
Up to a normalization, we recover the wave functions of the isolated ring
recalled in the previous subsection.
By integration in the ring we obtain:
\be
\int_0^l \D x\,|\psi_{(a)}(x)|^2 \APPROX{k\sim k_n^\pm}
\frac{1}{2k} \,\frac1\pi\frac{w^2/2l}{(k-k_n^\pm)^2+(w^2/2l)^2}
\,\leadto{w\to0}\, \frac{1}{2k}\delta(k - k_n^\pm)
=\delta(E - [k_n^\pm]^2)
\:,\ee
which is the correct DoS of the isolated ring.

\noindent$\bullet$ \mathversion{bold}$\theta=0$.\mathversion{normal}
If we now consider the degenerate case of zero flux, the resonance peaks
are in $k_n=2n\pi/l$. The wave function in the ring near the resonance is 
\be
\psi_{(a)}(x) \APPROX{k\sim k_n} \frac{1}{\sqrt{\pi k}}\, 
\frac{\I w/l}{k-k_n + \I w^2/l}\cos(2n\pi x/l)
\:.\ee
In this case the scattering state only reproduces the symmetric wave function
$\varphi^+_n(x)$ of the isolated ring. It is now clear that the integration
can not give the DoS of the isolated ring: indeed,
\be
\int_0^l \D x\,|\psi_{(a)}(x)|^2 \APPROX{k\sim k_n}
\frac{1}{2k}\, \frac1\pi\frac{w^2/l}{(k-k_n)^2+(w^2/l)^2}
\,\leadto{w\to0}\, \frac{1}{2k}\delta(k - k_n)
\ee
misses the degeneracy 2 of the eigenstates.

\vspace{0.25cm}

{\it In the cases studied above} the stationary scattering states,
computed in the limit of a graph weakly coupled to the leads ($w_\alpha\to0$), 
do not reproduce all the wave functions of the isolated graph and the 
{\it l.h.s. of the formula (\ref{SmiG}) can not be identified with the DoS 
of the graph}. In this sense the scattering states do not form a complete
basis to describe the Hilbert space of the graph.


\section{Remark on free graphs connected to one lead}

Note that for a free graph, {\it i.e.} with no potential and no magnetic
flux, perfectly connected ($w=1$) to only one lead, we can prove with
(\ref{rpd}) that 
\be
\delta^f(k^2) \APPROX{k\to0} 2k{\cal L}
\ee 
where
${\cal L}=\frac12\sum_{\alpha,\beta}a_\ab l_\ab$ is the total length of the
graph. The proof is obtained by analyzing the behaviours of the two spectral
determinants at small energy \cite{AkkComDesMonTex00}.
If the FSR fails, the total length ${\cal L}$ of the free graph is not 
encoded in the Weyl part of $\delta^f(k^2)$ but however ${\cal L}$ appears in 
the low energy behaviour of the phase shift.
For example if we consider the complete graph (figure \ref{fig:k5s}), we
have shown that $\delta^f_{\rm Weyl}=3k\ell$ and we can check on (\ref{pscg})
that the low energy behaviour ($k\to0$) is 
$\delta^f\simeq V(V-1)k\ell=2k{\cal L}$.


\section{Discussion}

We have shown that the well-known Friedel sum rule (FSR), a state counting 
method from the scattering properties, may be violated for certain graphs 
having degenerate spectrum. This has been demonstrated already at the 
level of the Weyl term of the DoS: we have studied several examples where
a discrepancy occurs between the Weyl term of the Dos of the isolated graph 
and the Weyl part of the derivative of the Friedel phase 
$\frac{1}{2\pi}\frac{\D\delta^f}{\D E} 
= \frac{1}{2\I\pi} \tr{\Sigma^\dagger\frac{\D\Sigma}{\D E}}$.

A way to understand the origin of the failure of the FSR is to
compare the quantities involved in the DoS and in the Friedel phase.
The density of states of an isolated graph can be obtained from the
spectral determinant $S(\gamma)=\prod_n(\gamma+E_n)$ by doing the
substitution $\gamma\to-E-\I0^+$. The spectral determinant is
proportional to the determinant of the matrix $M$ introduced above:
$S(\gamma)\propto\det M(\gamma)$ 
\cite{PasMon98,Pas98,AkkComDesMonTex00,Des00,Des00a,Des01}.
Adding a small imaginary part to the spectral parameter $\gamma$ produces
a resonance structure in $\partial_\gamma\ln S(\gamma)$, each peak having
a weight equal to the degeneracy of the state.
If we now consider the Friedel phase we note that the width of the resonances
are obtained by adding to the anti-hermitian matrix $M(-E)$ an hermitian
matrix $W^{\rm T}W$: the Friedel phase involves $\det(M(-E)+W^{\rm T}W)$. 
Comparing this latter determinant with the determinant $\det M(-E-\I0^+)$ 
involved in the density of states, it is not surprising that the Friedel 
phase does not produce the correct spectral weights since the ways the 
energies (zero of determinant) acquire an imaginary part is different in the 
two cases.

Another way to understand the failure of the FSR for graphs is the 
following. For a problem invariant under rotations in a $d$-dimensional space,
the essence of the FSR is to count the number of nodes of the wave function
in the angular channel crossing a $(d-1)$-dimensional sphere at infinity when 
the energy is varied. The number of states coincides with the number of nodes,
that is with the number of jumps of $\pi$ of the phase shift $\eta_l(E)$ of 
the partial wave of orbital momentum $l$.
On the other hand, a graph is connected to the external only through leads 
plugged at vertices. In a sense the Friedel phase counts the number of nodes 
of the wave function $\psi_k(x)$ that reach those vertices by varying $k$.
The failure of the FSR is caused when several nodes of $\psi_k(x)$ reach 
at the same energy the same vertex from different bonds issuing from this 
vertex (we can easily convinced ourselves of this remark by considering 
the ring connected to one lead studied above).

We can also provide a clear picture of the problem within the arc formulation
introduced in \cite{TexMon01} and recalled in section \ref{sec:Intro}. 
In the arc formulation, the wave function is described by a set of amplitudes.
Each arc $i$ is associated with a couple $A_i,B_i$. We gather the internal
amplitudes in a vector $A^{\rm int}$, the external in a vector $A^{\rm ext}$
and all amplitudes in a vector $A$. The internal amplitudes of the graph are 
related through the bond scattering matrix~:
$A^{\rm int}=R\,B^{\rm int}$. 
All amplitudes are also related with each other by the vertex scattering 
matrix~:
$B = Q\,A$. 
If we eliminate $B^{\rm int}$ we get~:
\bea
\tilde Q^{\rm T}\, A^{\rm ext} 
           &=& (R^\dagger-Q^{\rm int}) \, A^{\rm int} \\
B^{\rm ext}&=& \tilde Q\, A^{\rm int} + Q^{\rm ext}\,A^{\rm ext}
\,.\eea
In general $\det(R^\dagger-Q^{\rm int})\neq0$ whatever $k$ is and at all
energies of the continuous spectrum, the stationary scattering states are
the only solutions of the Schr\"odinger equation on the graph.
However, for certain graphs (in particular for those examined above),
there exists a discrete set of energies in the continuous spectrum for which 
$\det(R^\dagger-Q^{\rm int})=0$. This means that additionaly to the 
scattering states, we can construct at those particular energies solutions 
such that $A^{\rm ext}=B^{\rm ext}=0$ while the internal amplitudes satisfy
$(R^\dagger-Q^{\rm int})A^{\rm int}=0$ and $\tilde Q A^{\rm int}=0$. These 
two last equations describe a solution localized in the graph and that does
not communicate with the leads. The stationary scattering states give the 
solutions of the Schr\"odinger equation for the continuous spectrum
apart for a discrete set of energies where some additional states are 
localized in the graph and thus are not probed by scattering leading to
the failure of the state counting method from the scattering.

The study of the various examples of section \ref{sec:VoFSR} leads us to
make the following conjecture for the ability of $\delta^f$ to count the
states (at least at the level of the Weyl term): if there are degenerate
energies of degeneracies $d_n$, the Friedel phase $\delta^f$ counts
correctly the states of the system if $L\geq d_n$ leads are plugged at
vertices in such a way that the wave function can not vanish at the
positions of all these vertices at the same time.


\section*{Acknowledgements}

I am grateful to Alain Comtet, Jean Desbois and Gilles Montambaux for many 
stimulating discussions. 
I also acknowledge Markus B{\"u}ttiker for interesting remarks.


\begin{appendix}

\section{The stationary scattering states \label{app:SSS}}

In this appendix we recall briefly how the stationary scattering states are
constructed \cite{TexMon01}. 
We consider the stationary scattering state
$\psi^{(\alpha)}(x)$ of energy $E=k^2$ which describes a plane wave entering 
the graph from the lead connected at vertex $\alpha$ and being scattered by 
the graph into all leads.
On the lead connected to vertex $\mu$, the wave function is:
\be\label{wfl}
\psi^{(\alpha)}_{{\rm lead}\:\mu}(x) = 
\delta_{\mu\alpha} \EXP{-\I kx} + \Sigma_{\mu\alpha} \EXP{\I kx}
\:,\ee
where $x\in[0;+\infty[$.
The wave function on the internal bond $(\mb)$ of the graph, 
is related to the two linearly independent solutions
$f_\mb(x_\mb)$ and $f_\bm(x_\mb)$ of the differential equation
\be\label{Schroed}
\left( -\D_{x_\mb}^2 + V_{(\mb)}(x_\mb) - k^2 \right) f(x_\mb) = 0
\:\ee
for $x\in[0;l_\mb]$. The two solutions $f_\mb$ and $f_\bm$
satisfy the following boundary conditions at the edges of the
interval:
\be
\left\{ 
\begin{array}{l}
f_\mb(\mu)   = 1 \\ [0.1cm]
f_\mb(\beta) = 0
\end{array}\right.
\hspace{0.5cm} \mbox{and} \hspace{0.5cm}
\left\{ 
\begin{array}{l}
f_\bm(\mu)   = 0 \\ [0.1cm]
f_\bm(\beta) = 1
\end{array}\right.
\ee
where $f(\mu)\equiv f(x_\mb=0)$ and $f(\beta)\equiv f(x_\mb=l_\mb)$.
The stationary scattering state on the bond $(\mb)$ is:
\be\label{wfb}
\psi^{(\alpha)}_{(\mb)}(x_\mb) = 
\psi^{(\alpha)}_\mu\,f_\mb(x_\mb) + \psi^{(\alpha)}_\beta\,f_\bm(x_\mb)
\ee
which already satisfies the continuity condition (\ref{CL1}).
The relation between the functions $f_\mb$ and $f_\bm$ and the reflection 
and transmission coefficient characterizing the potential on the bond 
is established by computing the derivatives of $f_\mb$ and $f_\bm$ at 
the boundaries of the interval.
Imposing the ``current conservation'' (\ref{CL2}) then leads to the 
expressions (\ref{RES2},\ref{RES3},\ref{RES4}) that permit a systematic
construction of the scattering matrix \cite{TexMon01}.


\section{Friedel sum rule and Smith relation in one dimension\label{app:FSR1D}}

The one-dimensional case can be considered as a graph with one bond and 
two vertices and can therefore be described with 
the formalism presented in \cite{TexMon01} and this paper.
The FSR (\ref{RBU}) has been demonstrated in general terms in
\cite{DasMaBer69}, however it is interesting to give a rapid demonstration 
that follows the lines of the original one in three dimensions 
\cite{BetUhl37,Fri52,Hua63,LanLif66e}~; note also that it has been
demonstrated in \cite{AviBan85} that the Friedel phase in a one-dimensional
situation is related to the phase of the transmission amplitude (see
also \cite{TanBut99}).
We consider the one-dimensional hamiltonian $-\D_x^2+V(x)$ with $x\in\RR$ 
with a potential $V(x)$ being  concentrated in some region of 
the space.
We start by describing several possible basis of eigenstates caracterizing 
the scattering problem.

\noindent$\bullet$ The stationary scattering states of energy $E=k^2$  
related to the scattering matrix
\be
\tilde\Sigma = 
\left(
\begin{array}{cc}\tilde r&\tilde t'\\ \tilde t&\tilde r'\end{array}
\right)
\ee
are the state $\varphi^{(L)}(x)$ associated with a plane wave coming from
the left and $\varphi^{(R)}(x)$ for an incoming wave from the right.
The asymtotic behaviours of the left stationary scattering state are
$\varphi^{(L)}(x) = \EXP{\I kx}+\tilde r\,\EXP{-\I kx}$ for $x\to-\infty$
and $\varphi^{(L)}(x) = \tilde t\,\EXP{\I kx}$ for $x\to+\infty$.
The state $\varphi^{(R)}(x)$ involves similarly the coefficients $\tilde r'$ 
and $\tilde t'$. Note that
those states can be introduced even if the potential is not concentrated
in a finite interval provided that it decreases sufficiently rapidly at 
infinity.

\noindent$\bullet$ If the potential has a support $[x_1;x_2]$ we introduce
the stationary scattering states $\psi^{(L)}(x)$ and $\psi^{(R)}(x)$
related to the scattering matrix $\Sigma$: 
the left stationary scattering state behaves like 
$\psi^{(L)}(x) = \EXP{\I k(x-x_1)}+r\,\EXP{-\I k(x-x_1)}$ for $x\leq x_1$
and $\psi^{(L)}(x) = t\,\EXP{\I k(x-x_2)}$ for $x\geq x_2$. A similar
expression for $\psi^{(R)}(x)$ involves the coefficient $r'$ and $t'$:
$\psi^{(R)}(x) = t'\EXP{-\I k(x-x_1)}$ for $x\leq x_1$ and
$\psi^{(R)}(x) = \EXP{-\I k(x-x_2)}+r'\,\EXP{\I k(x-x_2)}$ for $x\geq x_2$.
The reflexions and transmissions defined in this way are naturally involved 
in transfer matrices, which makes one of the interest of this definition.

Comparing the two sets of eigenstates it is clear that
$\varphi^{(L)}(x)=\EXP{\I kx_1}\psi^{(L)}(x)$ and
$\varphi^{(R)}(x)=\EXP{-\I kx_2}\psi^{(R)}(x)$. The relations between the
coefficients of the two scattering matrices $\tilde\Sigma$ and $\Sigma$ are
then:
$r=\tilde r\,\EXP{-2\I kx_1}$, $r'=\tilde r'\,\EXP{2\I kx_2}$,
$t=\tilde t\,\EXP{\I k(x_2-x_1)}$ and
$t'=\tilde t'\,\EXP{\I k(x_2-x_1)}$. The relation between matrices reads:
$\Sigma={\cal U}\tilde\Sigma{\cal U}$ with 
${\cal U}={\rm diag}(\EXP{-\I kx_1},\EXP{\I kx_2})$.

\noindent$\bullet$
To derive the FSR we introduce the two eigenstates of energy $E=k^2$ 
labelled by the index $\sigma=1,2$:
\be
\Psi_{\sigma}(x) = [a_{\sigma,+}\,\theta(x)+a_{\sigma,-}\,\theta(-x)]\,
\sin(k|x|+\eta_\sigma(k^2)+\pi/2)
\hspace{0.5cm}\mbox{for}\hspace{0.5cm}|x|\to\infty
\:,\ee
where $\theta(x)$ is the Heaviside function.
If the potential is symmetric the two amplitudes $a_{\sigma,+}$ and
$a_{\sigma,-}$ are equal in modulus and $\sigma=1,2$ labels the symmetric
and antisymmetric states.
Let us establish the relation with the $2\times2$ scattering matrix
$\tilde\Sigma$.
We look for the relation between this basis of eigenstates and the 
first basis introduced above: let us write
$\Psi_\sigma(x) = \varphi^{(R)}(x)+C\,\varphi^{(L)}(x)$. Comparing 
their behaviours at $x\to+\infty$ we get
$\tilde r' + C\,\tilde t = \EXP{2\I\eta_\sigma}$
and at $x\to-\infty$:
$\tilde t' + C\,\tilde r = C\,\EXP{2\I\eta_\sigma}$.
These two equations show that 
$
\EXP{4\I\eta_\sigma}-(\tilde r+\tilde r')\EXP{2\I\eta_\sigma}
-\tilde t\tilde t'+\tilde r\tilde r'=0
$.
In other terms: 
\be\label{proof1}
\det(\tilde\Sigma-\EXP{2\I\eta_\sigma})=0
\:.\ee
Therefore $\EXP{2\I\eta_1}$ and $\EXP{2\I\eta_2}$ are the two eigenvalues of 
the scattering matrix $\tilde\Sigma$. To finish the proof of the FSR we 
consider that the system is in a large interval $[-R;+R]$ and impose Dirichlet 
boundary conditions.
The quantification condition for $\Psi_{\sigma}(x)$ reads 
$k_nR+\eta_\sigma(k_n^2)+\pi/2=n\pi$.
We introduce $\delta k_n=k_{n+1}-k_n$, therefore in the limit $R\to\infty$:
\be
\frac{1}{\delta k_n} 
\simeq \frac{R}{\pi} + \frac1\pi \frac{\D \eta_\sigma(k_n^2)}{\D k_n}
\ee
which is the density of modes in the channel $\sigma$. The term
$\frac{R}{\pi}$ is the density of modes in the absence of the potential:
$\frac{1}{\delta k^{(0)}_n}$. In the limit $R\to\infty$ the difference
of densities of modes $\frac{1}{\delta k_n}-\frac{1}{\delta k^{(0)}_n}$ 
remains finite. It follows that 
\be
\int_{-\infty}^{+\infty}\D x\,\left[\rho(x;E)-\rho_0(x;E)\right]
 = \frac1\pi \sum_{\sigma=1,2}\frac{\D \eta_\sigma(E)}{\D E}
\:\,\ee
where $\rho(x;E)=\bra{x}\delta(E-H)\ket{x}$ is the LDoS and $\rho_0(x;E)$ the 
LDoS in the absence of the potential.
Due to (\ref{proof1}) this equation can be rewritten
$
\int_{-\infty}^{+\infty}\D x\,\left[\rho(x;E)-\rho_0(x;E)\right]
= \frac{1}{2\I\pi}\tr{\tilde\Sigma^\dagger\frac{\D
\tilde\Sigma}{\D E}}
$. 
{\sc Qed}.

Next we would like to apply both the FSR and the Smith formula on a
simple example. We now consider a potential with support $[0;{\cal L}]$ that
vanishes elsewhere, a situation where it is meaningful to introduce $\Sigma$
(instead of $\tilde\Sigma$).

\noindent ({\it i}) The Smith formula (\ref{SmiG}) gives the DoS of 
the interval $[0;{\cal L}]$ 
\be\label{DoSG}
\int_0^{\cal L}\D x\,\rho(x;E) = 
\int_0^{\cal L} \D x\, 
\left(|\tilde\psi_E^{(L)}(x)|^2+|\tilde\psi_E^{(R)}(x)|^2\right)
= \frac{1}{2\I\pi}
\left(
  \tr{\Sigma^\dagger\frac{\D\Sigma}{\D E}} 
+ \frac{1}{4E}\tr{\Sigma-\Sigma^\dagger}\right)
\:.\ee
The two terms correspond to left ($L$) and right ($R$) stationary 
scattering states, which form a complete basis of eigenstates in one 
dimension.
Note that this relation has also been given in \cite{GasChrBut96} for 
the one-dimensional case.

\noindent ({\it ii}) On the other hand the FSR
\be\label{DMB}
\int_{-\infty}^{+\infty}\D x\,\left[\rho(x;E)-\rho_0(x;E)\right] =  
\frac{1}{2\I\pi} \left(
  \tr{\Sigma^\dagger\frac{\D\Sigma}{\D E}} 
 - \frac{\I \,{\cal L}}{\sqrt{E}}
\right)
= \frac{1}{2\I\pi}\tr{\tilde\Sigma^\dagger\frac{\D
\tilde\Sigma}{\D E}}
\:\ee
measures the variation of the DoS of the infinite line due to the presence 
of the potential in the interval $[0;{\cal L}]$. In particular (\ref{DMB}) 
is sensitive to the effect of the potential on the wave function at infinity
whereas (\ref{DoSG}) is a local quantity.

As an illustration, let us consider the extremely simple case of
a potential $\lambda\delta(x)$ on a line. The corresponding graph is a vertex
($V=1$, $B=0$ and $L=2$). Formulae (\ref{q1},\ref{q2},\ref{q3}) give
the scattering matrix
\be
\Sigma = \frac{2}{2+\I\lambda/k}
\left(\begin{array}{cc} 1&1\\1&1 \end{array}\right) - 1
\:.\ee
It is easy to check that (\ref{DoSG}) therefore vanishes
\be
\tr{\Sigma^\dagger\frac{\D\Sigma}{\D E}} 
+ \frac{1}{4E}\tr{\Sigma-\Sigma^\dagger}=0
\:,\ee
which is not surprising since the ``graph'' is only a point (the support of
the potential is an interval of measure $0$).
On the other hand the variation of the DoS of the infinite line can be 
computed either with the exact Green's function, known for this 
potential, or with (\ref{DMB}). The Green's function gives:
\be\label{vdid}
\int_{-\infty}^{+\infty}\D x\,\left[\rho(x;E)-\rho_0(x;E)\right] =  
\theta(-\lambda)\delta(E+\lambda^2/4) - \frac12\delta(E)
+ \theta(E)\frac{\lambda}{4\pi\sqrt{E}}\frac{1}{E+\lambda^2/4}
\:,\ee
where $\theta(E)$ is the Heaviside function. The first term is the
contribution of the bound state (that exists if $\lambda<0$). We can check 
that the total number of states is conserved: \\
$\int_{-\infty}^{+\infty}\D E
\int_{-\infty}^{+\infty}\D x\,\left[\rho(x;E)-\rho_0(x;E)\right]=0$.
The Friedel phase, obtained from the above scattering matrix is 
$\delta^f(E)=-\I\ln\det\Sigma=2\arctan(2k/\lambda)$ and we can therefore 
recover the expression (\ref{vdid}) using the FSR (\ref{DMB}).


\mathversion{bold}
\section{The spectrum of the complete graph $K_V$\label{app:GC}}
\mathversion{normal}

We give here the spectrum of the complete graph, which is made of $V$ 
vertices all connected among each other with bonds of same lengths $\ell$. 
The spectrum is easily extracted from the spectral determinant 
$S(-k^2)=\prod_{m=0}^\infty(E_m-k^2)$ which has been computed in 
\cite{AkkComDesMonTex00}:
\be
S(-k^2) \propto \left(\frac{\sin k\ell}{k}\right)^{\frac{V(V-3)}{2}}
\sin^2(k\ell/2) \:(\cos k\ell + \cos\varphi)^{V-1}
\ee
up to some inessential numerical factor. The parameters $\lambda_\alpha$ that
characterize the boundary condition (\ref{CL2}) are put to zero here. 
We have introduced $\cos\varphi=\frac{1}{V-1}$. The energies $E_m$ and the 
corresponding degeneracies $d_m$ are given in the following table:
\begin{center}
\begin{tabular}{|l|l|}
\hline
$k_m=\sqrt{E_m}$               & $\hspace{0.25cm}d_m$ \\
\hline
$k_0=0$                        & $1$   \\
$k_1=\frac{\pi-\varphi}{\ell}$ & $V-1$ \\
$k_2=\frac{\pi}{\ell}$         & $\frac{V(V-3)}{2}$ \\
$k_3=\frac{\pi+\varphi}{\ell}$ & $V-1$ \\
$k_4=\frac{2\pi}{\ell}$        & $2+\frac{V(V-3)}{2}$ \\
$\hspace{0.5cm}\vdots$         & $\hspace{0.5cm}\vdots$\\
\hline
\end{tabular}
\end{center}
It is obvious from the expression of the spectral determinant that the
spectrum is periodic in $k$ of period $2\pi/\ell$, that is:
$k_{m+4}=k_m+2\pi/\ell$ and $d_{m+4}=d_m$ (for $m>0$).

We next consider the corresponding eigenfunctions.
The eigenfunction on the bond $(\ab)$ is given by (\ref{wfb}):
\be
\psi_{(\ab)}(x) = \frac{1}{\sin kl_\ab}
\left(
  \psi_\alpha \sin k(l_\ab-x) + \psi_\beta \sin k x
\right)
\:.\ee
Imposing the conditions (\ref{CL2}) leads to the $V$ equations:
\be\label{cuc}
\sum_\beta M_\ab \psi_\beta = 0
\:.\ee
For a free graph $M$ is given by (\ref{fmM}) \cite{AvrSad91}. For the
complete graph we have:
\be\label{MKn}
M_{\ab}(-k^2) = \frac{\I}{\sin k\ell}
\left(\delta_\ab (V-1)\cos k\ell - a_\ab\right)
\ee
where the adjacency matrix is $a_\ab=1-\delta_\ab$.

\noindent$\bullet$ {\bf Zero Mode}. The wave function is constant on the
graph $\psi^{(0)}(x)=1/\sqrt{\cal L}$ where ${\cal L}=\frac{V(V-1)}{2}\ell$
is the total length of the graph.

\noindent$\bullet$ \mathversion{bold}$k=k_1$\mathversion{normal}.
All the matrix elements of $M$ are equal:
$M_\ab=-\frac{\I}{\sin\varphi}$. The equation (\ref{cuc}) has $V-1$
solutions labelled by $j=1,2,\cdots,V-1$. A possible basis is:
$\psi_\alpha^{(1,j)}=\delta_{\alpha,1} - \delta_{\alpha,j+1}$, up to a
normalization (this basis is not orthogonal).

\noindent$\bullet$ \mathversion{bold}$k=k_2$\mathversion{normal}.
At this energy the matrix $M$ is divergent and equation (\ref{cuc}) can not
give the eigenstates. They are obtained by considering the
equation $(1-RQ)A=0$ where $R$ and $Q$ are the matrices given by
(\ref{r1},\ref{q1},\ref{q2},\ref{q3}). $A$ is the vector gathering the
$2B$ amplitudes of the wave function (one for each arc) 
\cite{TexMon01,AkkComDesMonTex00}.
The system $(1-RQ)A=0$ has $V(V-3)/2$ solutions at $k=k_2$. To have an 
idea of the
structure of the solution let us consider $K_4$. We label by 1,2,3 and 4
the nodes, and we bring together the 6 components on the 6 bonds in a vector
$\Psi(x) = (\psi_{(12)}(x),\psi_{(13)}(x),\psi_{(14)}(x),\psi_{(23)}(x),
\psi_{(24)}(x),\psi_{(34)}(x))$.
We have for the first state, labelled $(2,1)$:
$\Psi^{(2,1)} = (1,0,-1,-1,0,1)\times\sin(\pi x/\ell)$,
and for the second eigenstate:
$\Psi^{(2,2)} = (0,1,-1,-1,1,0)\times\sin(\pi x/\ell)$.

\begin{figure}[!h]
\begin{center}
\includegraphics[scale=0.8]{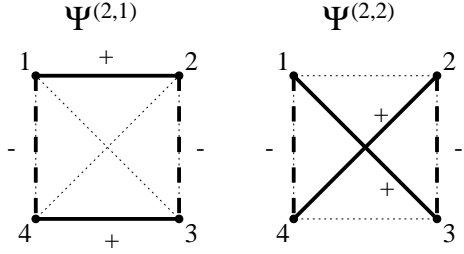}
\end{center}
\caption{The two eigenstates of $K_4$ of energy $k=k_2$. The dotted lines 
are the bonds and the large dots the vertices (labelled 1,2,3 and 4). 
The large continuous lines are put where the wave function is positive and the 
large dashed lines where it is negative. On the other bonds the wave function
vanishes.}
\label{fig:psi2K4}
\end{figure}

\noindent$\bullet$ \mathversion{bold}$k=k_3$\mathversion{normal}.
The matrix $M$ is the opposite as the one computed at $k=k_1$ and the 
wave functions on the nodes have the same value as for this latter energy.

\noindent$\bullet$ \mathversion{bold}$k=k_4$\mathversion{normal}.
The same problem occurs as for $k=k_2$. The system $(1-RQ)A=0$ has
$V(V-3)/2+2$ solutions corresponding to wave fonctions vanishing at all the 
nodes. Again we consider the graph 
$K_4$ and give the four degenerate states:
$\Psi^{(4,1)} = (0,0,0,-1,1,-1)\times\sin(2\pi x/\ell)$,
$\Psi^{(4,2)} = (0,-1,1,0,0,-1)\times\sin(2\pi x/\ell)$,
$\Psi^{(4,3)} = (-1,0,1,0,-1,0)\times\sin(2\pi x/\ell)$ and
$\Psi^{(4,4)} = (-1,1,0,-1,0,0)\times\sin(2\pi x/\ell)$.

\begin{figure}[!h]
\begin{center}
\includegraphics[scale=0.8]{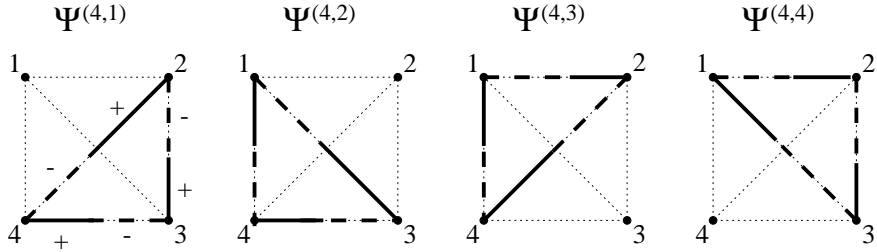}
\end{center}
\caption{The four eigenstates of energy $k=k_4$.}
\label{fig:psi4K4}
\end{figure}

\noindent$\bullet$ \mathversion{bold}$k=k_{m>4}$\mathversion{normal}.
The spectrum is periodic in $k$ with period $2\pi/\ell$. Then the value
of the wave functions at the nodes is the same at $k_m$ and $k_{m+4}$, only
the number of oscillations on the bonds change.

\end{appendix}



\end{document}